\documentclass[proof]{WileyASNA-v2}

\articletype{Article Type}%

\received{1 December 2018}
\revised{1 January 2019}
\accepted{1 January 2019}

\begin{document}

\title{Oblique magnetic fields and the role of frame dragging}

\author[1]{Vladim\'{\i}r Karas*}

\author[1]{Ond\v{r}ej Kop\'a\v{c}ek}

\author[2]{Devaky Kunneriath}

\author[1,3]{Tayebeh Tahamtan}

\authormark{V. Karas \textsc{et al}}

\address[1]{\orgdiv{Astronomical Institute}, \orgname{Czech Academy of Sciences}, \orgaddress{\state{Prague}, \country{Czech Republic}}}

\address[2]{\orgdiv{North American ALMA Science Center}, \orgname{National Radio Astronomy Observatory}, \orgaddress{\state{Charlottesville (VA)}, \country{USA}}}

\address[3]{\orgdiv{Institute of Theoretical Physics}, \orgname{Faculty of Mathematics and Physics, Charles University}, \orgaddress{\state{Prague}, \country{Czech Republic}}}

\corres{*Bo\v{c}n\'{\i} II 1401, CZ-14100 Prague, Czech Republic. \email{vladimir.karas@cuni.cz}}


\abstract{Magnetic null points can develop near the ergosphere boundary of a rotating black hole by the combined effects of strong gravitational field and the frame-dragging mechanism. The electric component does not vanish in the magnetic null and an efficient process or particle acceleration can occur. The situation is relevant for starving (low-accretion-rate, such as the Milky Way's supermassive black hole) nuclei of some galaxies which exhibit only episodic accretion events. The presence of the magnetic field field of an external origin is an important aspect. We propose that such conditions can develop when a magnetized neutron star approaches the supermassive black hole during late stages of its inspiral motion. The field lines of the neutron star dipole thread the black hole's event horizon and change rapidly their connectivity. We put in comparison the case of a dipole-type magnetic field of a sinking and orbiting star near a non-rotating black hole, and the near-horizon structure of an asymptotically uniform magnetic field of a distant source near a fast-rotating black hole. Although the two cases are qualitatively different from each other, they both develop magnetically neutral null points near the event horizon.}

\keywords{Black holes, Magnetic fields, Accretion disks, General relativity}



\jnlcitation{\cname{%
\author{Karas, V., et al.}} (\cyear{2019}), 
\ctitle{Oblique magnetic fields and the role of frame dragging}, \cjournal{Astronomische Nachrichten}, \cvol{2019;}.}

\maketitle

\footnotetext{\textbf{Abbreviations:} GR, general relativity; AGN, active galactic nucleus; SMBH, super-massive black hole.}

\section{Introduction}\label{sec1}
The present day astrophysics accepts the notion of black holes as a likely explanation for intriguing properties of super-massive dark
objects in cores of active galactic nuclei and quasars, as well compact stellar-mass components of X-ray binaries.
According to the classical general-relativity definition of a vacuum, asymptotically flat spacetime of an isolated black hole with 
a non-singular event horizon, black holes do not support their own, intrinsic magnetic field \citep{Misner:1974qy}. 
This can be understood in terms of black hole uniqueness theorems \citep{Wald:1984rg}, which limit the family of parameters of black-hole 
solutions \citep{Carter:1971zc}. The situation starts to be more complicated if the traditional assumptions are relaxed
\citep{Chrusciel:2012jk}. For astrophysical black holes it appear that only the mass and angular momentum are two relevant parameters, values of which need to
be constrained and measured by observation of electromagnetic light from nearby stars and gas, and the gravitational wave signal that can 
be released in the moment of black hole emergence and/or collision and coalescence with a secondary black hole or a neutron star 
\citep{2003ftpc.book...17R}. The only trivial exceptions are in principal related to the possibility of global charges on 
the black hole, although this does not seem to be an astrophysically plausible concern.

\begin{figure*}[tbh!]
\center
\includegraphics[scale=.37]{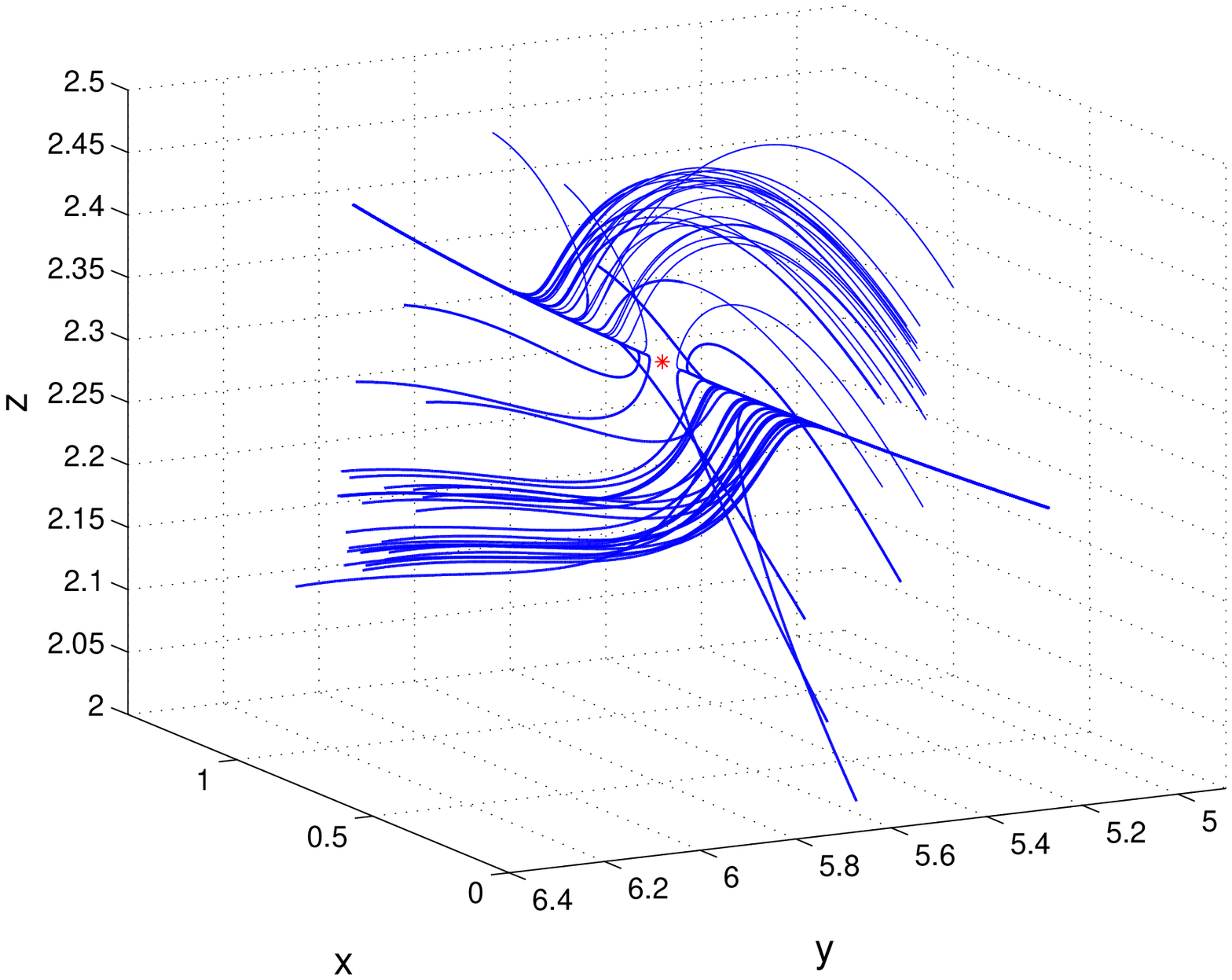}
\includegraphics[scale=.37]{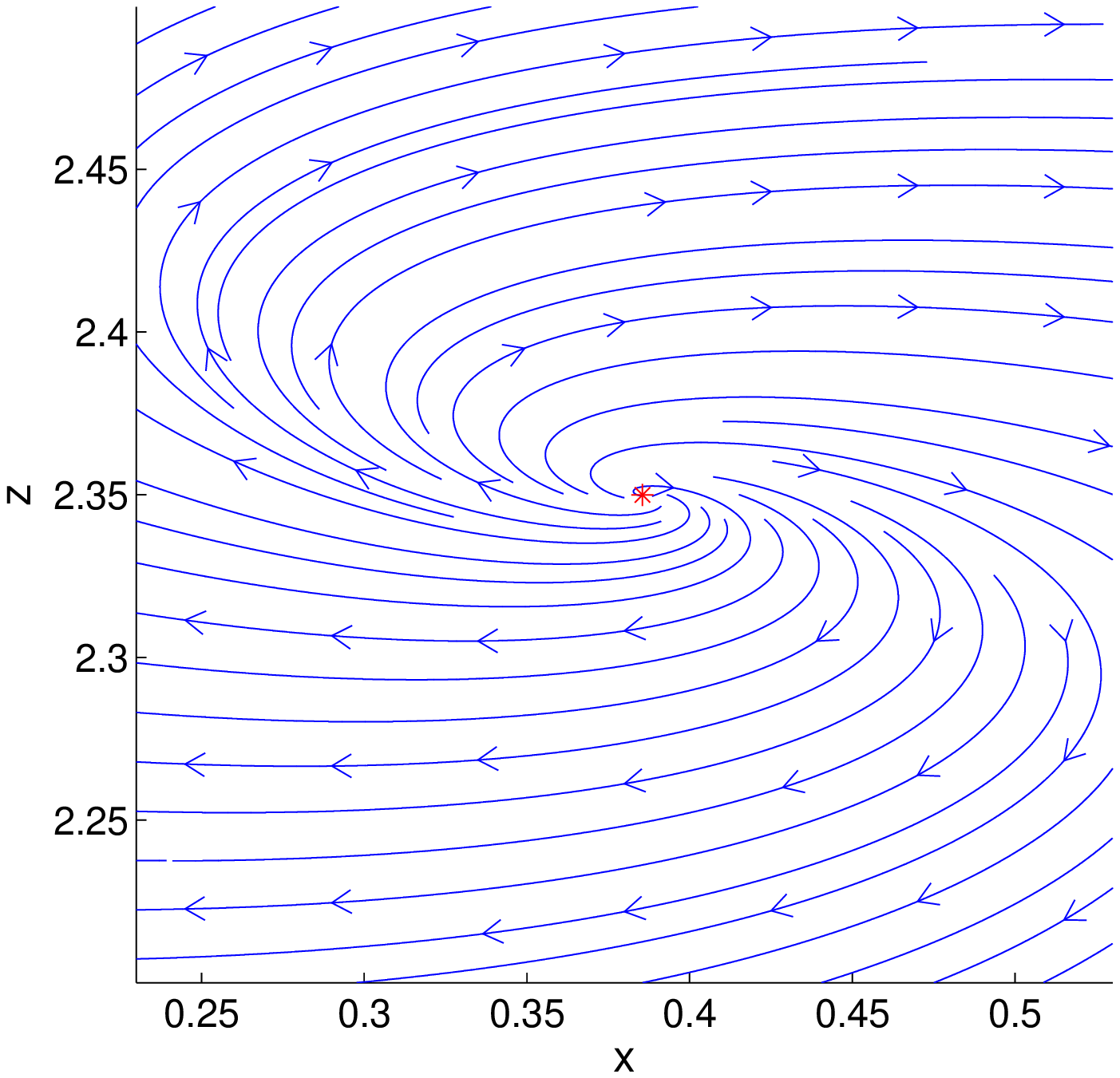}
\includegraphics[scale=.37]{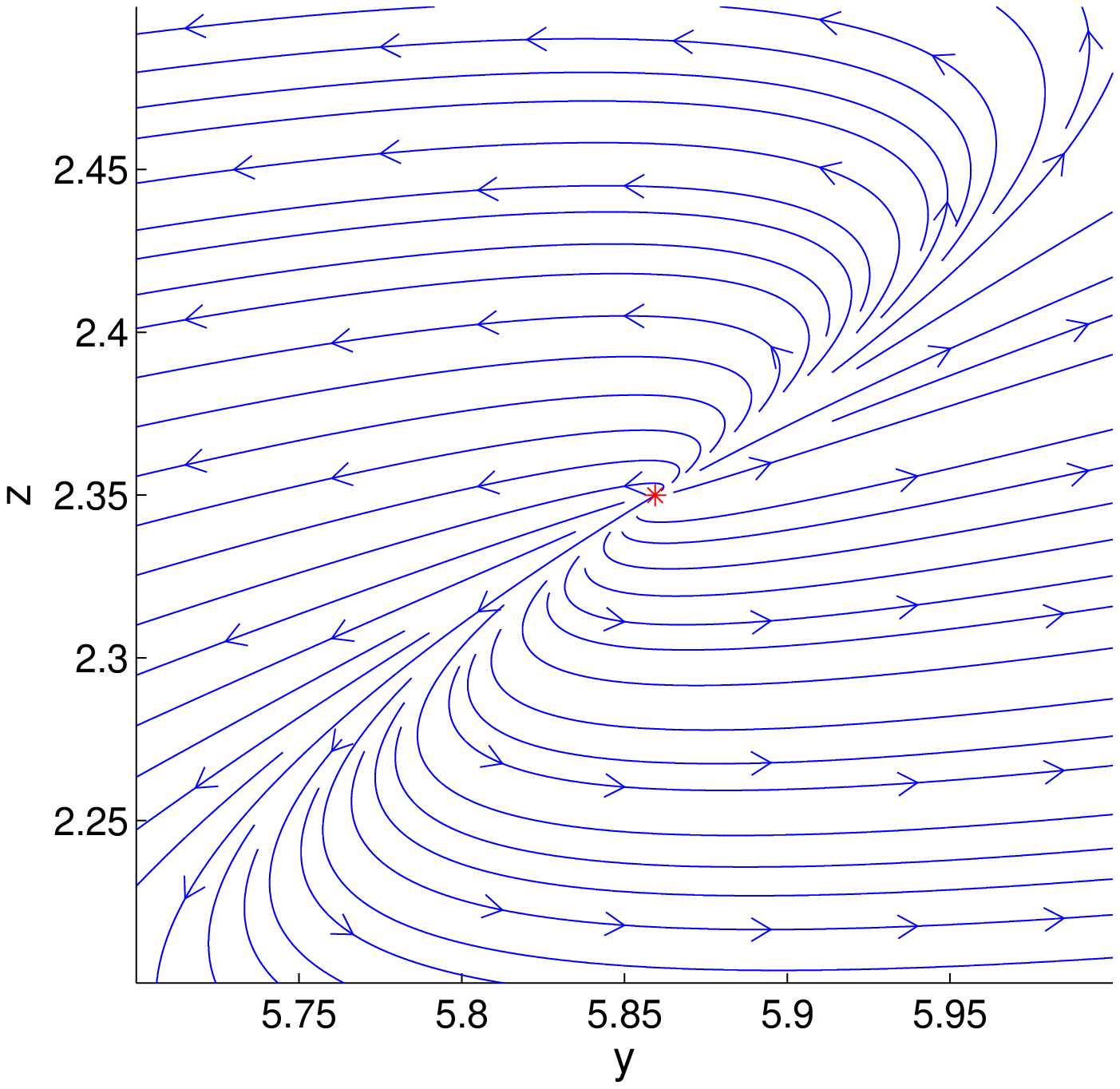}
\caption{Left panel: 3D projection of the magnetic field lines of a dipole near the null point (red mark). Middle and right panels: two-dimensional sections constructed across the null point in the Rindler approximation (figure adapted from \citeauthor{Kopacek:2018rkb} \citeyear{Kopacek:2018rkb}).}
	\label{NP_2D}
\end{figure*}

\section{Black hole immersed in an external magnetic field}\label{sec2}
In general theory of relativity, gravity shapes the space-time curvature, which then determines the way how particles move and how embedded
fields (the electromagnetic field in particular) bend and twist their lines of force. Vice versa, particles and fields acts as sources in the coupled
set of partial differential equations for the space-time metric,
\begin{equation} 
R_{\mu\nu}-\textstyle{\frac{1}{2}}Rg_{\mu\nu}=8\pi T_{\mu\nu};
\end{equation}
see, e.g.,  \citep{Misner:1974qy,Chandrasekhar1983}. On the right-hand side the source terms consist of contributions from both the 
material terms (particles and fluids) as well as fields. For the latter, electromagnetic field is clearly of astrophysical relevance,
\begin{equation}
T^{\alpha\beta}=\frac{1}{4\pi}\left(F^{\alpha\mu}F^\beta_\mu- \frac{1}{4}F^{\mu\nu}F_{\mu\nu}g^{\alpha\beta}\right),
\end{equation}
${T^{\mu\nu}}_{;\nu}=-F^{\mu\alpha}j_{\alpha}$, ${F^{\mu\nu}}_{;\nu}=4\pi j^\mu$, ${^\star F^{\mu\nu}}_{;\nu}=4\pi\mathcal{M}^\mu$
in the usual notation (the asterisk denotes a dual quantity). In realistic circumstances, however, the electromagnetic fields are not intense enough to curve 
the spacetime; instead, they just respond to gravity as a test field. Nonetheless, the effect of electromagnetic acceleration acting
on electrically charged particles can still be significant, and so it is important to understand the structure of these test fields. 

A number of cornerstone discovery papers dealt with the mathematical structure of electro-vacuum test field solutions on the background of classical 
black holes \citep[][and further references cited therein]{1973blho.conf..451R,Wald:1974np,King:1975tt,Chitre:1975sc,Petterson:1975sg,1977RSPSA.356..351P,Bicak:1980du,0143-0807-21-4-304}. Interestingly enough, some important aspects of the solutions 
have not yet been completely studied to date. Indeed, the above-mentioned early papers have demonstrated that the magnetic and electric lines of force are highly
distorted by the presence of the horizon and the gravito-magnetic action of a rotating black hole \citep{Punsly2009}, however, the potential astrophysical impact of these distortions remained to be seen. 

An interesting possibility has emerged: magnetic field line asymptotically {\it perpendicular} with respect to rotating (Kerr) black hole. Magnetic null points then emerge in the equatorial plane, where the magnetic components vanish even in the purely {\it electro-vacuum} configuration. In such a case, electrically charged particles can be accelerated very efficiently because, in the magnetic null point, the Larmor gyrations do not interfere with the acceleration by the electric component. The effect was examined in several papers \citep{Karas:2012mp,Karas:2014rka,2015JPhCS.600a2070K}; we found that a translation boost in combination with fast rotation of the black hole help to create the magnetic nulls. Here we further elaborate on the interesting features of such asymmetric (asymptotically oblique) magnetic fields in the near-horizon limit.

\begin{figure*}[tbh]
\center
\includegraphics[width=.31\textwidth]{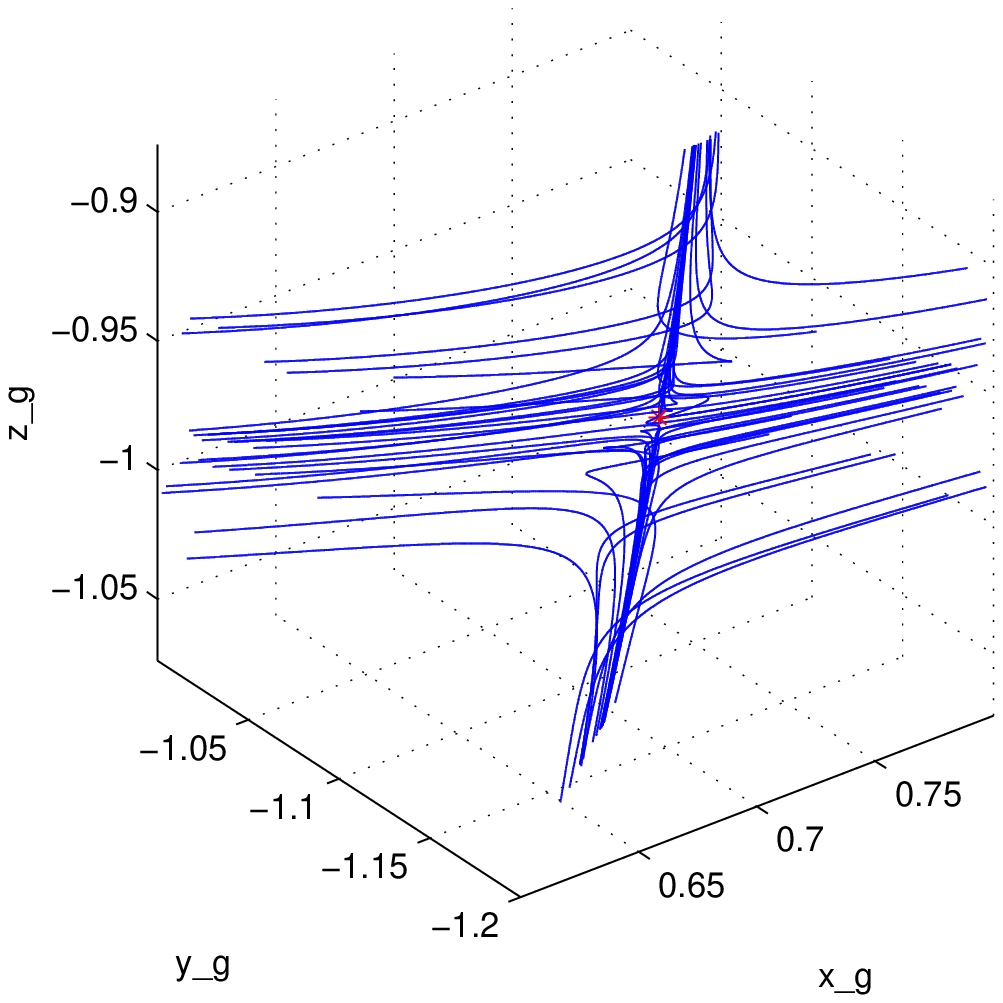}\hfill
\includegraphics[width=.28\textwidth]{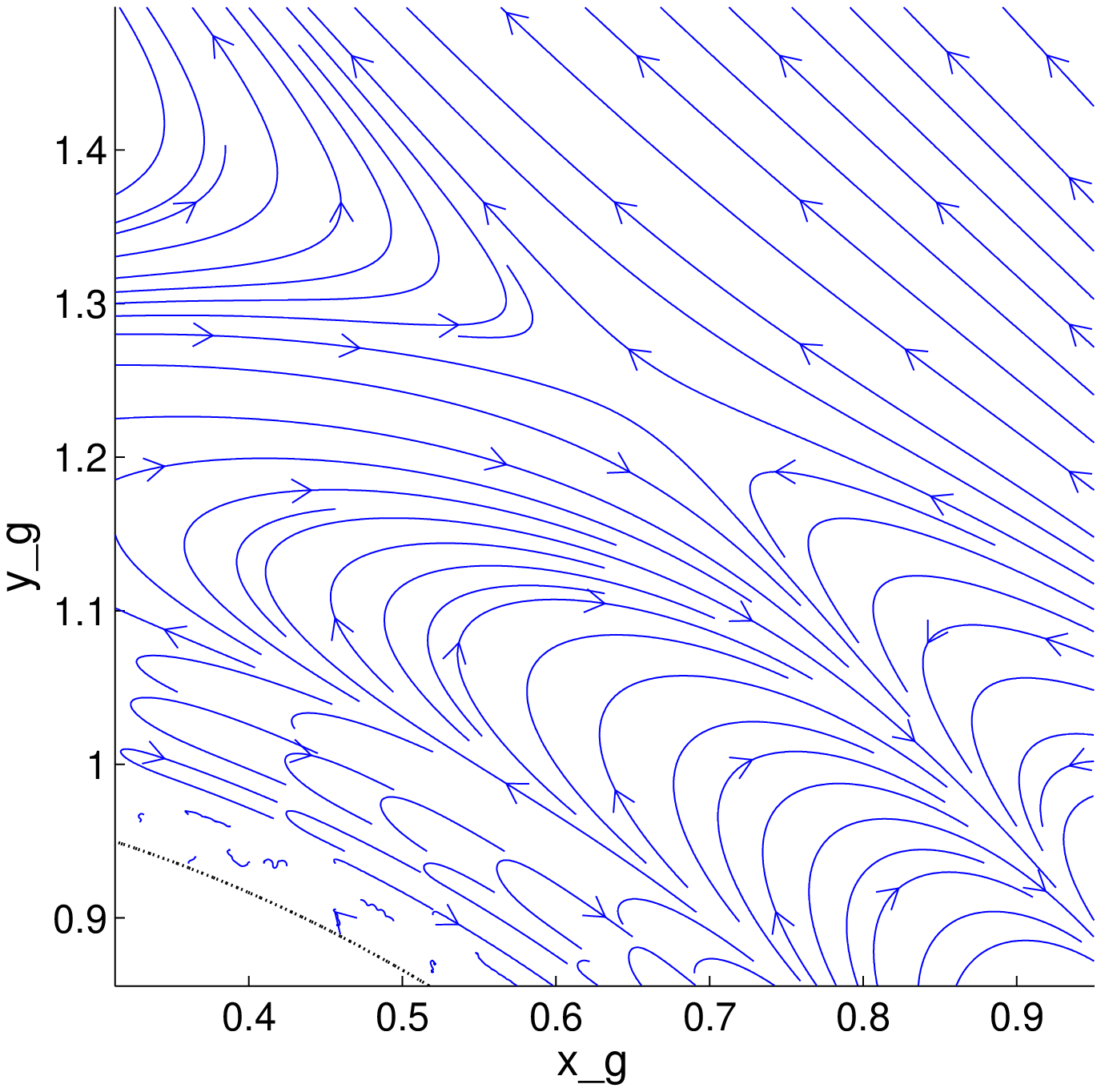}
\includegraphics[width=.35\textwidth]{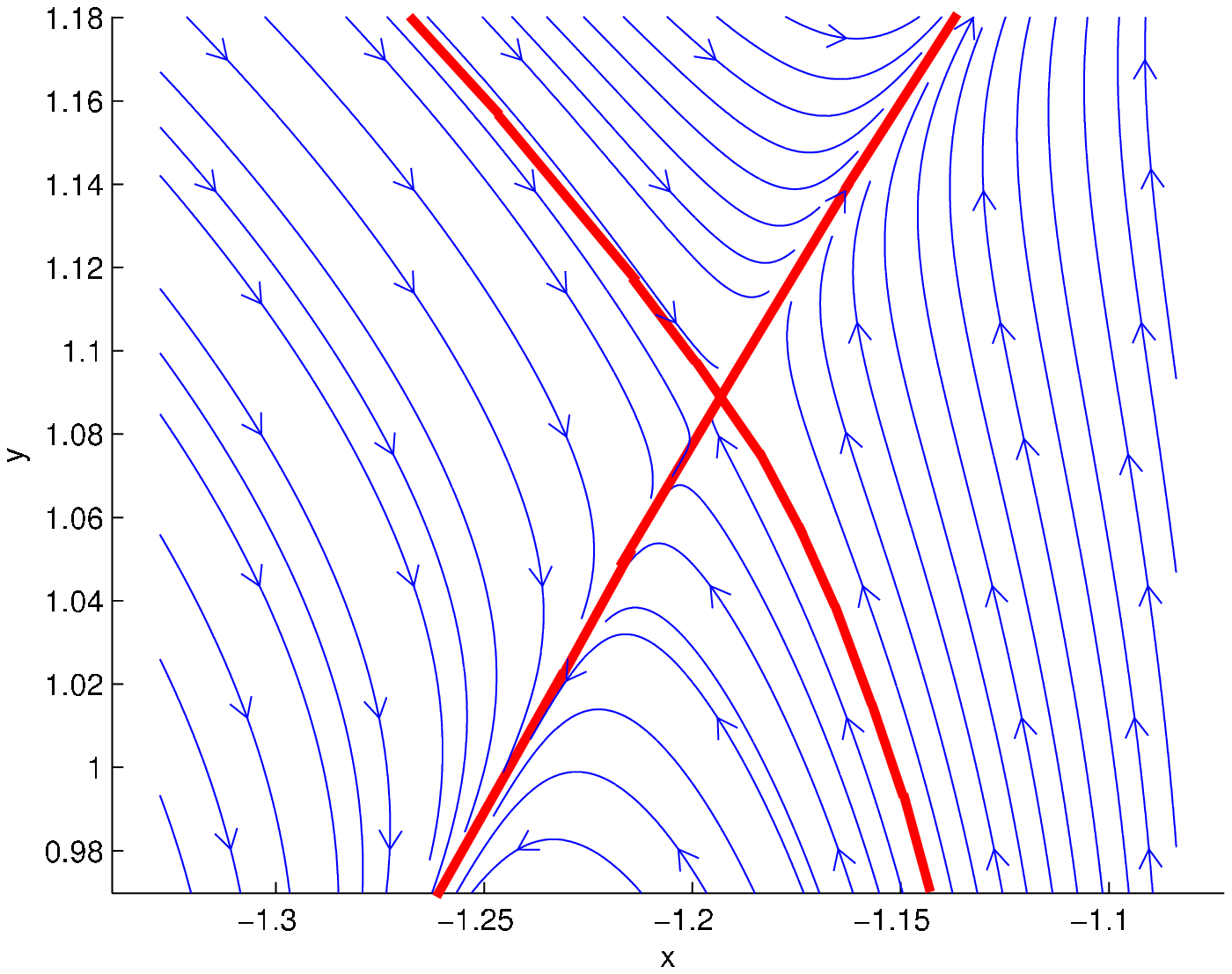}
\caption{Left panel: magnetic field lines of an asymptotically uniform field and the critical point (red cross) near a rotating black hole. Middle and right panels: detailed structure in of the two-dimensional sections. The black hole is centered in the origin; part of the horizon is seen in the bottom-left corner of the middle panel.The separatrix is formed by lines crossing the critical point in the right panel \citep[][and reference cited therein]{Karas:2017wre}.
\label{fig2}}
\end{figure*}

\section{Magnetic null points near SMBH horizon}
\subsection{Case of non-rotating black hole}
Supermassive black holes (SMBH) reside in centres of galactic nuclei, where they are surrounded by dense nuclear star clusters \citep[NSC; e.g.,][]{Schodel:2014gna,2016MNRAS.457.2122G}, including massive stars. In the course of stellar evolution, inevitably, a population of magnetized neutron stars emerges within NSC. This provides a plausible scenario, which brings the magnetic field of an orbiting and gradually sinking to the center neutron star down to the SMBH horizon \citep{DOrazio:2013ngp,Kopacek:2018rkb}.

We assume that the neutrons star's external magnetic field is approximated by a rotating dipole in vacuum \citep{Kovar:2010ty}. In order to derive the structure of this boosted rotating dipole interacting with the SMBH, we adopt the Rindler approximation \citep{Rindler:1966zz,1985PhRvD..32..848M}. This scheme is adequate to capture the structure of the field lines in the near-horizon limit and it allows us to include also the effects of linear translation velocity of the dipole, which in turn helps to create conditions necessary for the emergence of magnetic nulls in electro-vacuum, i.e., even without the contribution of a conducting plasma, which is usually assumed and indeed must be also present in the neutron star magnetosphere \citep{2006ASSL..340.....S}.

Rindler's near-horizon approximation ignores the effects of spacetime curvature, instead, it represents the black hole gravitational attraction locally (on scales $\ell\ll M$) by the equivalent acceleration. For our problem the Rindler approximation is a very useful tool, as it allows us to analyze analytically the electromagnetic field structure very close to the horizon, where the acceleration plays the dominant role.\footnote{Let us note that the structure of weak electromagnetic fields on the curved, Kerr-metric background can be solved in terms of infinite series expansion with no further approximation. In this respect the Rindler's approach gives useful insight, but it is not necessary.} Let us remind the reader the approximate metric is flat; it can thus written in Minkowski coordinates $(T,X,Y,Z)$. In the case of non-rotating black hole,
\begin{eqnarray}
\label{mink} 
ds^2 & = & -dT^2 + dX^2 + dY^2 + dZ^2\\
 & = & -\alpha^2 dt^2 + dx^2 + dy^2 +dz^2,
\end{eqnarray}
where $\alpha$ (lapse) relates the Rindler coordinates $(t,x,y,z)$ with the corresponding the spatial part of the Schwarzschild metric near the horizon \citep{1985PhRvD..32..848M},
\begin{equation}
ds^2  =  \frac{r}{r-2M}\,dr^2 + r^2\,d\Omega^2
  \simeq  g_h^{-2}d\alpha^2 + (2M)^2\,d\Omega^2;
\end{equation}
$g_h\simeq \alpha/z$ is the horizon surface gravity, $z\simeq 4M\sqrt{1/2M-r}$ is the proper distance from the horizon. 

The electric and magnetic field lines, for an observer moving at four-velocity $u^{\mu}$, can be written i terms of 
the electromagnetic field tensor $F^{\alpha\beta}$:
\begin{equation}
E^{\alpha}=F^{\alpha\mu}u_{\mu}, \qquad B^{\alpha}=\textstyle{\frac{1}{2}}\epsilon^{\alpha\mu\gamma\delta}F_{\gamma\delta}\,u_{\mu},
\label{fieldlines}
\end{equation}
respectively. Integrating three-dimensional projections of the electric and magnetic vectors in eq.\ (\ref{fieldlines}), either numerically or analytically, one can find lines of electric and magnetic force, respectively. We checked that analytical and numerical approaches lead consistently to the identical shape of field lines, as they indeed should \citep{Kopacek:2018rkb}. Then, for practical reasons of exploring the vast parameter space, we employed the numerical approach to identify the location of magnetic null points, where the $B$-field vanishes. Figure~\ref{NP_2D} shows a typical structure of the field lines of a rotating dipole in arbitrary (inclined) orientation and velocity. 

\begin{figure}[tbh]
\center
\includegraphics[width=.47\textwidth]{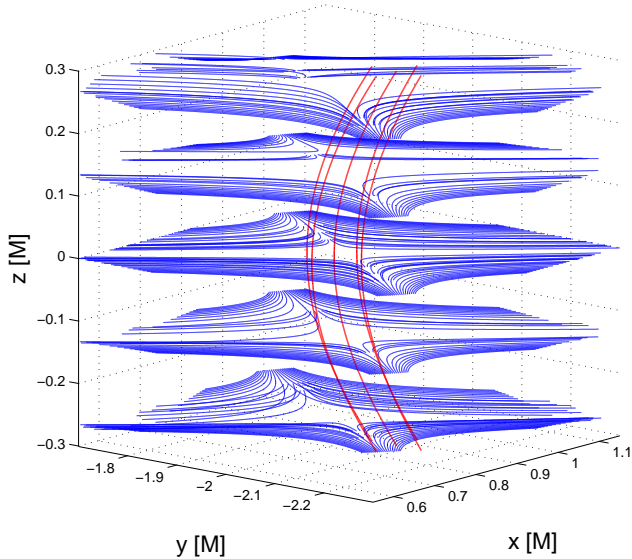}
\caption{The structure of electric and magnetic lines of force corresponding the homogeneous magnetic field perpendicular to the black hole rotation
axis $z$. In the equatorial plane ($z=0$) the magnetic lines (blue curves) reside in the plane, but they are progressively more distorted outside this plane; the X-type magnetic neutral point emerges. On the other hand, the induced electric lines (red curves) pass through the magnetic null point, thereby accelerating electrically charged
particles very efficiently \citep{Karas:2012mp}.
\label{fig3}}
\end{figure}

\subsection{Case of a rotating (Kerr) black hole}
When the black hole rotates, its gravitational effect of the surrounding matter and fields is different because of frame-dragging. In the context of producing the critical points, where the magnetic field disappears, the case with black hole rotation is, interestingly, both simpler and more complicated compared to the non-rotating case. On one side this case is simpler because the magnetic nulls occur even when the rotating black hole interacts with the homogeneous (asymptotically uniform) magnetic field; the frame dragging ensures that the field lines are dragged and distorted in a dramatic way near the ergosphere, where the space-time rotation becomes unbeatable. On the other side, the situation is more complicated because the space-time element contains cross-terms ($g_{\phi t}\neq0$), which means that the metric is stationary (not static) and the resulting gravito-magnetic action influences the field lines unlike the non-rotating (static) black hole. Fig.~\ref{fig2} compares the topology of field lines with the previous 
case.\footnote{Cartesian coordinates ($x_g, y_g, z_g$) correspond to the Boyer-Lindquist spheroidal coordinates ($r, \theta, \phi$) in this plot.}

Let us also notice a remarkable sequence of nested layers that develop near the black hole and are seen as one gradually approaches the event horizon. The orientation of the magnetic field reverses repetitively across these layers, which then correspond to current sheets in a non-vacuum (plasma filled) system. Moreover, a gravitomagnetically induced electric component can pass through the magnetic null point, thereby acting as an efficient acceleration mechanism for charged particles. This can be clearly illustrated in the case of a uniform magnetic field lying within the equatorial plane, where the electric lines rise above and below the plane; see Fig.\ \ref{fig3} for a
three-dimensional projection of the combined magnetic and electric lines in the vicinity of the magnetic null.

\section{Discussion}
We have demonstrated the emergence of magnetic neutral points in the purely electro-vacuum system, namely, in the vicinity of a magnetized black hole with fast rotation and the linear boost. The topology of magnetic lines is governed by the black hole strong gravity and the induced gravito-magnetic effects. The intricate structure of the lines of force is interesting by itself; moreover, the case of weak (test) electromagnetic field can be solved analytically in complete generality, although in this paper we started with the the simplest geometry of an asymptotically uniform (Wald-type) magnetic field that is homogeneous at spatial infinity, albeit the direction of the field can be arbitrary with respect to the black hole rotation axis, and the black hole can exhibit linear translation \citep[see][]{Kopacek:2018rkb} for a more complex case of a magnetic dipole near a black hole in Rindler's approximation.

Two aspects of the above-described electro-vacuum system should be mentioned. Regarding its efficiency to accelerate particles from the vicinity of a black hole, an electric component should be present in the magnetic null. While the Larmor gyrations in the null point vanish, electric field can accelerate charged particles. One can easily verify that, indeed, in our system the lines of electric force pass through the magnetic null (the origin of the electric component is due to gravito-magnetic effect, so the black hole rotation is required). In the case of the homogeneous magnetic field perpendicular to the black hole rotation axis in its equatorial plane, the non-vanishing electric component is directed \textit{perpendicularly} with respect to the equatorial plane (i.e. along the axis of symmetry). The escaping particles are thus also expected to be accelerated along the black hole rotation axis and they can thus form a stream outflowing in an escape corridor \citep{Kopacek:2018lgy}.

Let us note that the black hole system with an oblique (non-axisymmetric) magnetic field lacks any spatial symmetry. Therefore, the lines of force become entangled outside the magnetic null point, and the mechanism of their reconnection requires to apply the three-dimension (3D) formalism. Although this is beyond the scope of our present investigation, one can envisage the spine topology that governs the field lines outside the magnetic null \citep{Pontin:2011bk}. While in 2D the conditions for reconnection are confined to X-type null points (which in our system occur just in the equatorial plane and the magnetic field asymptotically perpendicular to the rotation axis), full 3D topology allows for even a richer structure; the reconnection process does not even have to be limited to the null points. The process of reconnection is then defined just by the changing connectivity of field lines. To this end, a non-vanishing component of parallel electric field is crucial. This is also why rotation of black hole became important in our scenario: it can generate the electric field by the gravito-magnetic interaction.

Unlike the usually assumed force-free approximation to model pulsar magnetospheres \citep[cf.][]{Petri:2013fsa,2019MNRAS.482.1942G} for a very recent account and references. Here we adopted the opposite limit and started from the electro-vacuum structure Finally, let us note that the traditional resistive MHD view of the reconnection does not apply to our discussion. The resistive mechanisms, however, are relevant in a sufficiently dense plasma. Instead, in highly diluted, hot environment, which is of particular relevance in space non-ideal reconnection effects are dominated by non-collisional processes and radiation cooling that are, typically, connected with Larmor gyrations \citep{Biskamp2000,Hakobyan:2018fwg}.

\section{Conclusions}\label{sec5}
We constructed the lines of magnetic force in the near-horizon region, where the electro-vacuum field of an external source interact with the black hole gravity. We considered the limit of an electromagnetic test field, which does not contribute as a source of the gravitational field in Einstein's equations, however, we allowed for the motion and/or rotation of the black hole. Magnetic null points can occur; the locus of vanishing magnetic intensity is a place where magnetic reconnection can be triggered once the plasma gets injected into the immediate vicinity of the magnetic neutral points, for example by the pair creation process.

\section*{Acknowledgments}
We thank the \fundingAgency{Czech Science Foundation}, grant ref.\ \fundingNumber{17-16287S}, titled ``Oscillations and Coherent Features in Accretion Disks around Compact Objects and Their Observational Signatures'', and the \fundingAgency{Czech Ministry of Education, Youth and Sports} COST program ref.\ \fundingNumber{LTC\,18058} to support international collaboration in relativistic astrophysics.

\bibliography{Karas}

\end{document}